\documentclass[twocolumn,aps,prb,showpacs]{revtex4}
\usepackage{graphics}
\begin{document}

\title{Is a ``homogeneous'' description of dynamic heterogeneities possible?}
\author{Grzegorz Szamel}
\affiliation{Department of Chemistry,
Colorado State University, Fort Collins, CO 80525}

\date{\today}


\begin{abstract}
We study the simplest model of dynamic heterogeneities in glass 
forming liquids: one-spin facilitated kinetic Ising model introduced
by Fredrickson and Andersen [G.H. Fredrickson and H.C. Andersen,
Phys. Rev. Lett. \textbf{53}, 1244 (1984); 
J. Chem. Phys. \textbf{83}, 5822 (1985)]. 
We show that the low-temperature,
long-time behavior of the density autocorrelation function
predicted by a scaling approach can be obtained from a 
self-consistent mode-coupling-like approximation.
\end{abstract}
\maketitle

Recently, a new general paradigm for dynamics of supercooled liquids has been 
proposed \cite{GarChan1}. It focuses directly on the spatial and temporal 
structure of dynamic heterogeneities,
\textit{i.e.} regions of higher-than-average mobility.
The essence of this new approach 
is a mapping of a microscopic model of a supercooled liquid
onto an effective model describing dynamics of 
the heterogeneities \cite{WhitBerGar}. 
It has been argued on physical grounds that very simple effective models can 
capture the most important features of the heterogeneities' dynamics. 
In the initial work \cite{GarChan1} two 
models were used: a one-dimensional one-spin 
facilitated kinetic Ising model, \textit{i.e.}~the Fredrickson-Andersen (FA)
model \cite{FA}, and 
its asymmetric version, \textit{i.e.}~the East model \cite{East}. 
In the subsequent work \cite{GarChan2}, 
more complicated (and more realistic) models
were used. However, it is the simplest model (\textit{i.e} the 
FA model) that is most often used \cite{BerGar1,BerGar2} 
to illustrate the main points of
the new paradigm and to contrast it 
with other approaches to the glass transition problem
like the mode-coupling theory \cite{Goetze} and the landscape 
paradigm \cite{landscape}.

The simple models describing dynamic heterogeneities have trivial
static properties. However, they exhibit interesting
dynamics. Hence, there arises a new problem: how to study the dynamics
of the effective models. One should notice that solving this problem will be, 
most likely, considerably simpler than solving the original,
microscopic model of a supercooled liquid: the effective models
used so far are lattice-based and thus have a greatly reduced number of
degrees of freedom; even more importantly, these models 
usually involve a small parameter (\textit{i.e.}~density of 
heterogeneities (mobile regions)) which can facilitate theoretical analysis.

The original work \cite{GarChan1} concentrated on the analysis of the
structure of trajectory spaces of the FA and East models; it was
argued that the glass transition coincides with an entropy crisis
in trajectory space rather than in configuration space. This is an
ingenious suggestion because it focuses directly on the system's dynamics and
thus it should be applicable to other glass-like systems with 
trivial equilibrium properties \cite{Renner, SS}. 

In Ref. \cite{GarChan1}, 
in order to get explicit results for dynamics of the FA and East models,
the random walk theory was
combined with scaling arguments. Subsequently \cite{WhitBerGar}
a dynamic field-theoretical model has been derived 
and analyzed using the renormalization group (RG) methods. It was
argued that the $d$-dimensional FA model belongs to the same universality
class as the field-theoretical model. RG analysis showed that,
for $d>2$, there is no finite temperature phase transition and
the low temperature scaling properties are determined by the directed
percolation critical point.

In our opinion, the simplicity of the effective models
suggests that it might be possible to analyze their dynamics
using well-known methods of non-equilibrium 
statistical mechanics. Such an approach was utilized by
Fredrickson and Andersen \cite{FA}: the
time-dependent density correlation function was
expressed in terms of a memory function; the latter function was then
analyzed using an exact diagrammatic expansion \cite{Pitts}. The resulting
self-consistent equation for the density correlation function
predicted a power law dependence of the relaxation time on the 
density of heterogeneities. However, neither Ref. \cite{FA} nor later
work on the FA model \cite{ReitJack} reproduced correct
scaling exponent for the relaxation time. This was, perhaps, the reason
for the recent claim \cite{BerGar2} that no ``homogeneous'', 
mean-field \cite{comment}
approach is able to reproduce scaling predictions for the FA model.

In this note we present a simple theory that does reproduce previously
predicted scaling behavior \cite{review} 
of the FA model density correlation function.
The theory relies upon a self-consistent factorization 
approximation of a pair-density 
correlation function into a product of two density correlations.
This approximation is analogous to one used in the
mode-coupling theory \cite{Goetze}; the important difference is that
we use the factorization approximation for pair-density
involving non-nearest-neighbors. We argue that while this makes 
the structure of the theory somewhat complicated, 
it is necessary from a physical point of view since the dynamics  
of nearest-neighbor up-spins are drastically different from that of 
individual spins. 

The theory presented here shows that it might be possible
to use established ``homogeneous'' methods to analyze dynamics 
of effective models of heterogeneities.

The FA model consists of a chain of $N$ non-interacting Ising spins in
a unit magnetic field.
We use density variables $n_i$ ($n_i =1$ corresponds to $i$th spin 
pointing up; physically, it corresponds to a more mobile region
located near site $i$). In terms of $n_i$ the Hamiltonian is 
$H = \sum_i n_i$. The average density is 
$\left<n_i\right> \equiv c = 1/(1+\exp(1/T))$ where $T$ is the temperature.  

We consider the normalized 
time-dependent density correlation function, $C_{i,j}(t)$,
\begin{equation}\label{denscf}
C_{i,j}(t) = \left<\delta n_i \exp\left(\Omega t\right) \delta n_j\right>
/\left<\left(\delta n_i\right)^2\right>
\end{equation}
where $\delta n_i$ is the deviation of the density at site $i$ from its
equilibrium average, $\delta n_i = n_i -c$; note that $C_{i,j}(t=0) = 
\delta_{i,j}$. In Eq. (\ref{denscf}) 
$\Omega$ is the evolution operator; time evolution consists of single
spin-flip dynamics with rate that is proportional to the number of
nearest neighbors in the spin up state. Explicitly, 
\begin{equation}\label{O}
\Omega = - \sum_{i=1}^N m_i \left(1-S_i\right) w_i
\end{equation} 
where $S_i$ is the spin-flip operator, $S_i n_i = -n_i +1$,
$m_i$ is the number of nearest neighbor up-spins, $m_i = n_{i+1} + n_{i-1}$,
and $w_i$ is the factor maintaining detailed balance,
$w_i = c + n_i (1-2c)$.

Some earlier simulational studies \cite{Harr} 
and recent computational and theoretical works 
\cite{GarChan1,BerGar1,BerGar2} used the so-called persistence function 
to characterize dynamics of the FA model. The persistence function
is defined as a fraction of spins that remain unflipped after a given
time. While this function is easier to determine from simulations
than the density correlation function,
it is somewhat awkward to analyze theoretically because it is defined
via a time-nonlocal constraint. 

The starting point of our analysis is the memory function representation
of the density correlations function,
\begin{eqnarray}\label{mfrep}\nonumber
&&\sum_j\left(\delta_{i,j} + M^{1irr}_{i,j}(z)
\left<\delta n_j m_j \delta n_j\right>^{-1} \right)
\left(z C_{j,k}(z) - \delta_{j,k}\right) 
\\  &&=
\left<\delta n_i \Omega \delta n_i\right> 
\left<\delta n_i \delta n_i\right>^{-1} C_{i,k}(z)
\end{eqnarray}
Here $C_{i,j}(z)$ is the Laplace transform of $C_{i,j}(t)$,
$C_{i,j}(z) = \int_0^{\infty} dt\, \exp(-zt) C_{i,j}(t)$,
and $M^{1irr}_{i,j}(z)$ is the Laplace transform of the 
irreducible \cite{irr1,irr2}
memory function. The memory function can be expressed in terms of the
pair-density correlation function involving nearest neighbor spins,
\begin{equation}\label{mf2spin}
M^{1irr}_{i,j}(t) = A_{i,j}(t) + A_{i-1,j}(t) + A_{i,j-1}(t)
+ A_{i-1,j-1}(t).
\end{equation}
Note that if the $M^{1irr}_{i,j}$ is neglected,
the density correlation function $C_{i,j}$ is diagonal; motion of 
the isolated up-spins is possible only due to the coupling to 
pair-densities. 

The pair density correlation function $A_{i,j}(t)$ evolves with 
1-spin irreducible
evolution operator $\Omega^{1irr}$,
\begin{equation}\label{2spin}
A_{i,j}(t) = \left<\delta n_i \delta n_{i+1} \exp\left(\Omega^{1irr}t\right)
\delta n_j \delta n_{j+1}\right>
\end{equation}
where 
\begin{equation}\label{O1irr}
\Omega^{1irr} = \hat{Q}_1 \left[ \Omega +
\sum_i \left. m_i \delta n_i \right> 
\left<\delta n_i m_i \delta n_i\right>^{-1}
\left<\delta n_i m_i \right. \right] \hat{Q}_1. 
\end{equation}
In Eq. (\ref{O1irr}) $\hat{Q}_1=1-\hat{P}_1$, and 
$\hat{P}_1$ is a projection operator on the one-spin density subspace,
\begin{equation}\label{P1}
\hat{P}_1 = \sum_i \left. \delta n_i \right> 
\left<\left(\delta n_i\right)^2\right>^{-1} \left<\delta n_i \right. 
\end{equation}

It should be noted at this point that the standard sequence of 
approximations leading to a mode-coupling theory \cite{Goetze} 
would start from Eq. (\ref{2spin}). 
Briefly, the pair-density correlation function 
would be factorized into a product of two density correlations, and the
irreducible operator $\Omega^{1irr}$ would be 
replaced by $\Omega$ \cite{MCTKaw}.
However, a closer investigation shows that the time dependence of 
$A_{i,j}$ is very different from that of the product
of $C_{i,j}$s. In particular, if in an equation of motion for $A_{i,j}$
one neglects the memory function term, one gets a (lattice) diffusion equation.
This is in contrast to Eq. (\ref{mfrep}) that, upon neglecting
the memory function term, leads to a site-diagonal density correlation 
function.

The strategy followed in this note is different: we propose to continue
using the memory function approach until we generate a function that
can be expressed in terms of pair-densities involving non-nearest
neighbors, in terms of \textit{e.g.}~$\delta n_i \delta n_{i+l}$, $l\ge 2$. 
With this goal in mind, we derive a memory function representation for the
pair-density correlation function, $A_{i,j}$. It turns out that the
2-spin irreducible memory function that enters the equation of motion
for $A_{i,j}$ can be expressed in terms of the triple-density
correlation function, $B_{i,j}$,
\begin{equation}\label{2mf3spin}
M^{2irr}_{i,j}(t) = B_{i,j}(t) + B_{i-1,j}(t) + B_{i,j-1}(t)
+ B_{i-1,j-1}(t),
\end{equation}
where $B_{i,j}$ is defined as
\begin{equation}\label{3spin}
B_{i,j}(t) = \left<\delta n_i \delta n_{i+1} \delta n_{i+2}
\exp\left(\Omega^{2irr}t\right)
\delta n_j \delta n_{j+1} \delta n_{j+2} \right>,
\end{equation}
and the 2nd order irreducible evolution operator $\Omega^{2irr}$ is 
given by a formula analogous to (\ref{O1irr}). It should be noted
that $\Omega^{2irr}$ involves operators $\hat{Q}_2$
projecting on the space orthogonal to the density and to the pair-density
involving \textit{nearest neighbor} sites, $\hat{Q}_2=1-\hat{P}_1-\hat{P}_2$.
Here $\hat{P}_1$ is given by Eq. (\ref{P1}) and $\hat{P}_2$ is defined as
\begin{equation}\label{P2}
\hat{P}_2 = \sum_i \left. \delta n_i\delta n_{i+1} \right> 
\left<\left(\delta n_i\delta n_{i+1}\right)^2\right>^{-1} 
\left<\delta n_i\delta n_{i+1}. \right.
\end{equation}

Next, we derive the memory function representation for $B_{i,j}$.
Its irreducible memory function is given by the following 
expression
\begin{eqnarray}\label{3mf4spin}\nonumber
&& M^{3irr}_{i,j}(t) = \left<\delta n_i \delta n_{i+1} \delta n_{i+2}
\left(m_{i+2} + m_{i}\right) \hat{Q}_3 
\right. \\  && 
\exp\left(\Omega^{3irr}t\right) 
\left. \hat{Q}_3
\left(m_{j+2} + m_{j}\right) 
\delta n_j \delta n_{j+1} \delta n_{j+2} \right>,\;\;\;
\end{eqnarray}
where $\hat{Q}_3=1-\hat{P}_1-\hat{P}_2-\hat{P}_3$,  
and $\hat{P}_3$ is defined as
\begin{eqnarray}\label{P3}\nonumber
\hat{P}_3 &=& \sum_i \left. \delta n_i\delta n_{i+1}\delta n_{i+2} \right> 
\left<\left(\delta n_i\delta n_{i+1}\delta n_{i+2}\right)^2\right>^{-1} \\ 
&& \left<\delta n_i\delta n_{i+1}\delta n_{i+2}. \right. 
\end{eqnarray}
Finally, the 3rd order
irreducible evolution operator $\Omega^{3irr}$ in 
Eq. (\ref{3mf4spin}) is 
given by a formula analogous to (\ref{O1irr}).

At this point we note that $M^{3irr}_{i,j}$ is a correlation
function of a quantity
that has a non-zero
overlap with the pair-density involving non-nearest neighbor sites:
\begin{eqnarray}\label{overlap}\nonumber
&& \left<\delta n_i \delta n_{i+2} \hat{Q}_3
\left(m_{i+2} + m_{i}\right) 
\delta n_i \delta n_{i+1} \delta n_{i+2}\right> \\ && 
= 2 \left(c(1-c)\right)^3.
\end{eqnarray}
Thus, it is at this stage that we propose to follow the usual sequence of
approximations \cite{Goetze}. As a result we get the following
expression for the memory function $M^{3irr}_{i,j}$ in terms
of the density correlations $C_{i,j}$s:
\begin{equation}\label{3mf4spinmct}\nonumber
M^{3irr}_{i,j}(t) \approx 
4 \left(c(1-c)\right)^4 \ C_{i,j}(t) C_{i+2,j+2}(t).
\end{equation}

The final self-consistent equation for the density correlation function
is easiest to present in terms of the Laplace-Fourier
transform, $C(q;z)$,
\begin{equation}\label{LF}
C(q;z) = \sum_k \exp(-iqk) C_{k,0}(z).
\end{equation} 
It reads
\begin{widetext}
\begin{equation}\label{self-cons}
C(q;z) = 
\frac{1}{\displaystyle 
z+\frac{2c}{\displaystyle 1+
\frac{2(1-c)\cos^2(q/2)}{\displaystyle z+2(1-c)\sin^2(q/2) +
\frac{2c}{\displaystyle 
1+\frac{2(1-c)\cos^2(q/2)}{\displaystyle z+2(1-c)\sin^2(q/2) + 
\frac{2}{\displaystyle 1+ 2c(1-c) \phi_2(q;z)}}}}}},
\end{equation}
\end{widetext}
where $\phi_2(q;z)$ is the Fourier-Laplace transform of 
$C_{i,j}(t) C_{i+2,j+2}(t)$.

To investigate the low temperature (\textit{i.e.}~low $c$) behavior
of the density correlation function we consider the scaling
function $\tilde{C}(q^*;z^*)$,
\begin{eqnarray}\label{Cscaling}\nonumber
\tilde{C}(q^*;z^*) &=& \lim_{c\rightarrow 0}
c^3 C(q^*c;z^*c^3) \\ &=& \int_0^{\infty} d t^* \exp(-z^*t^*)
\tilde{C}(q^*;t^*)
\end{eqnarray}
where $\tilde{C}(q^*;t^*) = \lim_{c\rightarrow 0} C(q^*c;t^*c^{-3})$. 
It can be easily
shown that in the limit $c\rightarrow 0$ the self-consistent 
equation (\ref{self-cons})
reduces to the following $c$-independent equation for the scaling function 
$\tilde{C}(q^*;z^*)$:
\begin{equation}\label{scalself-cons}
\tilde{C}(q^*;z^*) = \frac{1}{\displaystyle 
z^* + q^{*2}/2 + 1/\tilde{\phi}_2(q^*;z^*)}.
\end{equation}
Here $\tilde{\phi}_2(q^*;z^*)$ is the Laplace transform of 
$\tilde{\phi}_2(q^*;t^*)$,
\begin{eqnarray}\label{scalphi}\nonumber
\tilde{\phi}_2(q^*;t^*) &=& \lim_{c\rightarrow 0} 
c^{-1} \phi_2(q^*c;t^*c^{-3}) \\ &=& 
\int_{-\infty}^{\infty} \frac{dk^*}{2\pi} \tilde{C}(k^*;t^*)
\tilde{C}(q^*-k^*;t^*)
\end{eqnarray}

The scaling (\ref{Cscaling}) and $c$-independence of Eq. (\ref{scalself-cons})
shows that, in the low $c$ limit, up-spins (\textit{i.e.}~mobile regions) 
undergo diffusive motion with a diffusion constant $D\propto c$; 
up-spin diffusion is cut-off at long times by a longest relaxation
time $\tau\propto c^{-3}$. These results agree with known low $c$
scaling predictions \cite{review}.

In order to get quantitative results for the diffusion constant and
the relaxation time one would have to solve the scaling equation 
(\ref{scalself-cons}) combined with Eq. (\ref{scalphi}). 
The numerical solution of Eqs. (\ref{scalself-cons}-\ref{scalphi})
and a comparison of the resulting density correlation function 
with computer simulations are left for a future study.
Here we note that a single iteration of Eqs. 
(\ref{scalself-cons}-\ref{scalphi}) 
starting from a zeroth order approximation 
$\tilde{C}^{(0)}(q^*;z^*) = 1/\left(z^* + a q^{*2}/2 + b\right)$
leads to 
\begin{equation}\label{1iter}
\tilde{C}^{(1)}(q^*;z^*) = \frac{1}{z^* +  q^{*2}/2 + 
\left(4 a z^* + 8 a b + a^2 q^{*2}\right)^{1/2}}.
\end{equation}
Whereas the zeroth order approximation $\tilde{C}^{(0)}$ has a simple pole,
the first order approximation $\tilde{C}^{(1)}$ 
has a pole and a branch cut.
The simplest self-consistency condition that allows one to estimate
$a$ and $b$ (and thus the diffusion constant and the longest relaxation time, 
respectively) is to identify the pole in $\tilde{C}^{(0)}$
and the one in $\tilde{C}^{(1)}$. This leads to 
the following approximate results for the diffusion constant \cite{comment2} 
and the
longest relaxation time:  $D = 0.4 c$, 
$\tau = 0.3125 c^{-3}$.

One should note that the structure of Eq. (\ref{scalself-cons}) is similar
to that of the self-consistent equation of the mode-coupling theory
of the glass transition \cite{Goetze}. The very important difference is the
$q^{*2}/2$ term in the denominator. It is this term that cuts off the 
ergodicity-breaking transition that is generic for mode-coupling theories.

In summary, we have shown 
here that a simple self-consistent, mode-coupling-like
approximation reproduces known scaling predictions for the density 
correlation function of the FA model. It would be of interest to investigate
whether a similar approach can be applied to more complicated models like 
the East model or the Kob-Andersen lattice gas model \cite{KA},
which recently have attracted renewed interest \cite{GarChan1,Biroli}.

\textit{Note added:} After this paper had been submitted a preprint
presenting an RG approach to the one-dimensional FA model appeared 
\cite{WhitGar}.

Support by
NSF Grant No. CHE-0111152 is gratefully acknowledged.


\begin{thebibliography}{99}
\bibitem{comment2} It is has been claimed \cite{ReitJack} 
that the exact low density result for the 
diffusion constant is $D=c/2$. We believe that a more sophisticated 
self-consistent approximation may reproduce the exact value.
This point is left for a future study.
\end{thebibliography}
\end{document}